# Topologically Protected Ferroelectric Domain Wall Memory with Large Readout Current

Wenda Yang, Guo Tian[*], Hua Fan, Yue Zhao, Hongying Chen, Luyong Zhang, Yadong Wang, Zhen Fan[*], Zhipeng Hou, Deyang Chen, Jinwei Gao, Min Zeng, Xubing Lu, Minghui Qin, Xingsen Gao[*], and Jun-Ming Liu

Wenda Yang, Guo Tian, Hongying Chen, Luyong Zhang, Yadong Wang, Zhen Fan, Zhipeng Hou, Deyang Chen, Jinwei Gao, Min Zeng, Xubing Lu, Minghui Qin, Xingsen Gao, Jun-Ming Liu

Guangdong Provincial Key Laboratory of Quantum Engineering and Quantum Materials, Institute for Advanced Materials, South China Academy of Advanced Optoelectronics

South China Normal University

Guangzhou 510006, China

Email: X. S. Gao, xingsengao@scnu.edu.cn; Z. Fan, fanzhen@m.scnu.edu.cn; G. Tian, guotian@m.scnu.edu.cn

Hua Fan, Yue Zhao

The Department of Physics

Southern University of Science and Technology

Shenzhen 518000, China

Jun-Ming Liu

Laboratory of Solid-State Microstructures

Nanjing University

Nanjing 210093, China

**Abstract**

The discovery and precise manipulation of atomic-size conductive ferroelectric domain defects, such as geometrically confined walls, offer new opportunities for a wide range of prospective electronic devices, and the so-called walltronics is emerging consequently. Here we demonstrate the highly stable and fatigue-resistant nonvolatile ferroelectric memory device based on deterministic creation and erasure of conductive domain wall geometrically confined inside a topological domain structure. By introducing a pair of delicately designed co-axial electrodes onto the epitaxial $BiFeO_3$ film, one can easily create quadrant center topological polar domain structure. More importantly, a reversible switching of such center topological domain structure between the convergent state with highly conductive confined wall and the divergent state with insulating confined wall can be realized, hence resulting in an apparent resistance change with a large On/Off ratio $> 10^4$ and a technically preferred readout current (up to 40 nA). Owing to the topological robustness of the center domain structure, the device exhibits the excellent restoration repeatability over $10^6$ cycles and a long retention over 12 days ($> 10^6$ s). This work demonstrates a good example for implementing the exotic polar topologies in high-performance nanoscale devices, and would spur more interest in exploring the rich emerging applications of these exotic topological states.

**Introduction**

Ferroelectric domain walls, a sort of tunable atomic sharp homo-interfaces, have attracted widespread attention as an arena for exploring a wide range of emerging physical properties (e.g. enhanced conductivity, giant photo-voltages, and magnetoresistance) as well as immense application potentials [1-8]. For instance, highly conductive charged domain walls (CDWs) have been discovered in numerous initially insulating ferroelectric materials such as $BiFeO_3$ (BFO) [9-10], $BaTiO_3$ [11], $Pb(Zr,Ti)O_3$ [12], $LiNbO_3$ [13], and $YMnO_3$ [14]. These atomically sharp conductive channels can be created, eliminated, or displaced by external stimuli, constituting the basis of various energy-efficient and configurable nanoelectronic devices (i.e. domain-wall electronics) including nonvolatile memories [15-17] and memristors [18] among others. Moreover, these wall devices possess an appealing potential to be scaled down to ultra-small dimension mainly due to the atomic sharp nature of the conductive domain walls [17].

Recent studies did show some prototypes of domain-wall electronic devices (e.g. memories) based on the nonvolatile creation and erasure of CDWs and the nondestructive current readout [15-18]. However, such devices usually face the obstacle of unstable restoration process due to the difficulty in deterministic manipulation of the conduction channels in domain walls [10]. For instance, a CDW unit might not appear at the same location after an erasing and rewriting cycle, owing to the complexity of domain switching [19]. Moreover, the conductivity of a CDW unit may be greatly modulated by local distortions (e.g. bending) or defects, adding to the difficulty of deterministic control of domain wall conduction states [20].

In recent years, there has been a fast development in discovering novel complex polar topological objects (*e.g.* flux-closure, vortex, center states) that content conductive domain walls, likely offering new opportunities for deterministic manipulation of the domain wall conductions on the other hand [21-31]. For instance, the improved stability and switching repeatability of conduction states for CDWs confined in center topological domains have been found in well-prepared BFO nanoislands, benefiting from the effects of topological protection and geometric confinement [28]. More recently, robust electrically programmable quasi-one-dimension conduction channels were realized in the cores of center or vortex states confined in the BFO nanoislands [29]. These observations well exemplified that topologically protected and geometrically confined conductive wall units may be highly preferred candidates for emergent memory devices with excellent controllability and stability.

Although the reported topological protected conducting channels confined in ferroelectric nanoislands show promising potentials for high performance memories, it is still rather challenging to develop practical solid-state devices with patterned electrodes based on these small nanoislands. Therefore, an extensive application of these spatially confined and controllable topological states and conducting channels for realistic operation becomes an issue. Specifically, the long-term stability and high restoration repeatability in such devices are yet to be demonstrated, and in fact many proposed conceptual devices in this category face such obstacles to be practically utilized. On the other hand, a major but much less-addressed technical issue is the low readout current level of such nanoisland-based devices. Surprisingly, for many cases, such current is usually rather small (~1 nA), which is insufficient to drive a high-speed readout circuit according to the Johnson–Nyquist limit [10, 32, 33]. Therefore, it is of great interest to develop an alternative strategy for overcoming or avoiding these obstacles.

One of the best strategies, if applicable, would be to fabricate the devices directly from the well-prepared thin films, allowing integration with advanced microelectronic technologies. For example, one may design a set of electrode-patterned units within which a topologically protected central domain with a well-defined conductive channel can be developed, exhibiting the long-term stability, good restoration repeatability, and relatively large readout current. Here, we demonstrate such a prototype domain-wall memory unit based on highly conductive and confined quadrant center topological domain states. These structures are generated via a delicately designed co-planer electrode geometry by means of simple microfabrication methods from the well-prepared BFO thin films, as presented in the Methods section. The as-generated center-type domains can be easily created and reversibly switched between the divergent state that contains poorly conductive 71° tail-to-tail CDW (T-T CDW) and the convergent state that contains highly conductive 71° head-to-head CDW (H-H CDW). Such a memory unit demonstrates well repeatable resistive switching behavior with high On/Off ratio ($10^4$), and in particular the head-to-head polarization configuration is well preserved without interruption throughout the whole H-H CDW, leading to large readout current up to 40 nA, a highly preferred value for reliable reading operation. Furthermore, owing to the topological protection of both the T-T and H-H CDWs, the device exhibits the long retention (up to $10^6$ s under test) and high endurance (up to $2\times10^6$ cycles, the longest testing time), providing a new pathway towards high-performance topologically protected domain wall memories that are easily scaled-down and integrated.

**Creation of quadrant topological domains**

The epitaxial BFO thin films are deposited on the (001)-oriented SrTiO$_3$ (STO) substrates using pulsed laser deposition (PLD), and all the as-gown BFO films exhibit a mosaic domain structure, which is a pre-requisite and will be further utilized for creating quadrant center domains containing conductive DW (details of the preparation given in Methods). The rhombohedral BFO phase and its epitaxy are demonstrated by the X-ray diffraction (XRD) plus the reciprocal space mapping (RSM) (see Supplementary Figure. 1). To fabricate the device, an Au layer is deposited on the BFO film surface via thermal evaporation circularly shaped using the atomic force microscope (AFM) tip via the so-called AFM lithography to scrap the Au film in the central region. Thus, an array of circularly exposed BFO regions surrounded respectively by the circular Au electrodes is developed (seen in Figure. 1a). The fabrication details can be referred to the Methods section and Supplementary Figure. 2. It should be mentioned that the as-exposed BFO regions do exhibit piezoelectric properties as good as the as-prepared thin films with the mosaic domains (see Supplementary Figure. 3).

Such well-fabricated circular pattern enables further creation of center topological domains with the assistance of conductive AFM (C-AFM) tip. In sequence, by placing a stationary C-AFM tip on the center of the circular exposed BFO region, the tip and the surrounding Au electrode form the co-axial electrode-pair, which can generate radially distributed electric field in the exposed BFO region (see Figure. 1b). This is expected to radially align the in-plane ferroelectric polarization, so that a quadrant center-type topological domain, as restricted by the BFO crystal symmetry, can be generated after the electric field removal. It should be mentioned that such a center-type domain structure is highly robust once it is generated at room temperature.

To determine the local domain structure of the circularly exposed region, the vector piezoresponse force microscopy (PFM) is conducted upon the sample rotation of 0° and 90° (in fact at many angles) with respect to a reference axis, following previous method [27,29], which can largely determine the projection of local polarization vectors along respectively the [001], [010] and [100] axes of the BFO lattice. The out-of-plane (vertical) and in-plane (lateral) PFM images shown in Figures. 1c and 1d illustrate the created center domain structure and corresponding current maps probed by applying electric fields of ± 6.0 V between the tip and the surrounding Au electrode. For details, the bright and dark contrasts in the lateral PFM images indicate the opposite polarization components perpendicular to the cantilever direction. Given that the BFO crystal symmetry allows only eight possible [111] polarization directions, in combination with the PFM results, we can qualitatively reconstruct the local distribution of polarization structure using a MATLAB program, as shown in the in-plane projection of local polarization vector map (see the bottom row in Figures. 1c and 1d). The in-plane vector map clearly indicates that an electric pulse of +6 V (applied on the Au electrode) via the AFM tip is sufficient to create a quadrant center-convergent domain pattern (with upward vertical polarization), wherein all the in-plane polarizations are pointing inward to the center, and thus the cross-shaped head-to-head (H-H) charged domain wall (CDW) is generated. In reverse, an electric pulse of -6 V can produce a quadrant center-divergent domain, with the in-plane polarization pointing outward from the center. The cross-shaped tail-to-tail (T-T) CDW is thus generated.

It is also interesting to see that an electric biasing can simultaneously switch the vertical polarization of the exposed BFO region, as reflected by the contrast change in vertical PFM phase image (see the third row in Fig. 1c and d). This vertical polarization switching can also be demonstrated by the typical piezoresponse hysteresis loops, based on which the coercive voltages of about ± 4.8 V are revealed, analogous to those of bare BFO film on a bottom electrode (see Supplementary Figure. 4). The simultaneous reversal of both the in-plane and vertical components indicates that the center domain switching is a completely 180º reversible switching, which is more likely a collective switching behavior that is a character of the topologically-protected domain structure.

Subsequently, the local electric current profile can be mapped using the C-AFM mode under a sampling bias of 2.5 V. For the initial state with mosaic domain pattern, the exposed BFO surface exhibits very low conductive current (< 1 pA, close to the noise level), in contrast to the highly conductive current obtained on the electrode region (see Fig. S5). After applying a bias of +6.0 V to form a center-convergent domain, one can clearly see the apparent hallmark of the cross-shaped high conductive current pattern originating from the H-H CDW in the center-convergent state (see both Figure. 1c and Supplementary Figure. 5) [33]. Conversely, for the center-divergent domain triggered by - 6.0 V bias voltage, the cross-shaped current pattern disappears in the corresponding C-AFM map, as the T-T CDW shows very low conductivity that is comparable to domain interior. The reversible switching between the different domain wall conduction states shows very good repeatability, which is also likely the consequence of the topological protection property.

This observation conforms well with previous observation of conductive center domain state in BFO nanoislands [19, 28], while the total conduction current level in this work is one order larger in magnitude (>10 nA) compared with those observed in nanoislands (~1 nA) [28]. Such large conduction with the H-H CDW can be interpreted by the accumulation of charge carriers at the CDW with unbalanced net bound charges, which can greatly enhance the electric potential and thus attract electrons to screen the positive net bound charges. The bound charges can also bend the energy band inside the CDWs, which lowers the conduction band below the Fermi level, thus consequent to the metallic conducting behavior and significant enhancement in conductivity [9, 11]. This metallic conduction behavior can be further confirmed by the current-temperature ($I$-$T$) curves measured in CDWs (see Supplementary Figure. 6). It is seen that the current reduces with rising temperature, confirming the metallic conduction behavior [9, 16]. Unlike the conventional CDW that can be easily distorted by local perturbation, thus losing part of their conductivity [20], the topologically protected CDW is rather stable with very few distortions, which is favorite for maintaining their high conductivity.

**Resistive switching properties**

Based on the distinguished conductance switching between the convergent and divergent center states, one can propose a prototype of nonvolatile domain-wall memory device, and its configuration and operation can be given schematically in Figure. 2a. It is noted that the C-AFM tip is fixed at the center of exposed BFO region and serves as a central electrode, and the bias voltage is applied on the surrounding circular Au electrode. The characteristic of such DW memory is analogous to a highly repeatable resistive switching behavior, which can be well demonstrated by a quasi-static current-voltage (*I-V*) measurement at a sweeping bias range within ±7 V for 1,000 switching cycles. As shown in Figure. 2b, all the *I-V* curves exhibit rather symmetric hysteretic behavior. One can see that once the bias voltage increases to ~ 4.5 V, the readout current suddenly jumps from a low level (<1 pA) to a rather high level (~ 40 nA), and then increases linearly with further increasing bias. The sudden jump in the conductive current synchronizes well with the formation of highly conductive center-convergent state, consistent with the observation shown in Figure. 1c. The onset voltages (set and reset voltages) coincide with the coercive voltages shown in Supplementary Figure. 4. After the formation of center-convergent state with H-H CDWs, the conductive current increases linearly likely due to the metallic nature of the H-H CDW, as confirmed by the temperature-dependent C-AFM measurement mentioned above. The as-such created metallic channel can maintain its conductivity as the bias sweeps towards negative voltage, suggesting its non-volatility. It is also noted in the low bias range (within ±2 V), the current level becomes very small and the *I-V* curve no longer follows the linear behavior probably due to the presence of an interfacial insulating gap or wedge domain state close to electrodes [9,11]. Such interfacial insulating state is also analogous to a bipolar selector for the resistance random access memory cell, which is promising for suppressing the unwanted sneak current paths in a crossbar-array integrated electroresistance memory [34].

Besides, the resistance switching behavior as illustrated from the *I-V* curve gives rise to a large On/Off ratio of more than four orders in magnitude at a relatively low readout voltage (2.5 V). The nonvolatile data readout is also indicated by performing minor *I-V* curves at sweeping bias range of ±3.0 V, indicating a large On/Off resistance ratio over $10^4$ (see Figure.

2c). More interestingly, the device exhibits highly repeatable *I-V* characteristics. One can see that the hysteretic *I-V* curves follow almost the same trace of the first cycle after 1,000 cycles. From the cycling data (see Figure. 2a), we can extract the relative distribution of readout current (Log(I)) for both LRS and HRS recorded at a reading bias of 2.5V, as well as those of set and reset voltages, as shown Figure. 2d. It is seen that all the above four parameters exhibit very narrow distribution range, confirming the ultrahigh repeatability for data restoration of the device. This also well exemplified the high repeatability for the conduction state switching of CDWs confined in center topological domains, indicating the important role of the topological protection effect, an advantageous feature for the high-performance nonvolatile memory with nondestructive current readout.

Furthermore, the effect of lateral dimension of devices on the resistive switching properties is also examined, as shown in Supplementary Figure. 7. Similar resistive switching behavior (*i.e. I-V* curves) can be observed for different characteristic dimensions (i.e. the diameters of the exposed BFO regions) ranging from 200 nm to 500 nm, as reflected by the sharp resistance change that corresponds to the switching between convergent and divergent center states (see Figure. 2e). More interestingly, the dimension shrinkage can lead to an apparent reduction in onset resistance switching voltage (set voltage) as well as an increase in readout current at a fixed reading bias ($V_r$), *e.g.* the readout current in LRS increases from 8 nA for lateral size 500 nm to 15 nA for 200 nm at a $V_r$ of 2.5 V, and increases from 17 nA for 500 nm to 40 nA for 200 nm at a $V_r$ of 4.0 V, almost following an exponential law (see Figure. 2f). In practical device, these large reading current levels are essential for a high signal to noise level and for driving a fast operation [32, 33]. The above observation also indicates that with the dimension downscaling, the performance of devices can be further improved (enhancement in readout current and reduction in set voltage). It was theoretically predicted that the quadrant center domain can remain stable at an ultra-small dimension (~16 nm) [28], implying the huge potential of such device for developing ultrahigh density memory while maintaining their good device switching characteristics.

**Performances of domain wall memory device**

To further understand the memory performance of the as-generated device, we examine the memory switching window, retention, endurance, and fast switching dynamics of a randomly selected device, as shown in Figure. 3. Repeated testing of other devices shows the similar consequence while quantitative difference may exist. The memory switching hysteresis for this selected device can be found in Figure. 3a, and the data are collected using a sequence of electric pulses, and the resulting conductive states are readout under a fixed $V_r$ of 2.5 V. The readout current loop exhibits a rather square hysteresis with abrupt jumps at around ±4.5 V, showing a well-established memory window that is analogous to its corresponding piezoresponse hysteresis loop.

Besides, the stability of the memory device is further confirmed by the retention testing, as shown in Figure. 3b, and the data show very stable conductive states for both the LRS and HRS over long retention duration up to $10^6$ s (i.e.12 days, the longest testing time of this work and much longer duration is expectable). The high stability is further verified by a high temperature retention test, which demonstrates that topological domain structures and the corresponding resistance states can maintain rather good stability at a temperature of 150°C for $10^4$ seconds except some slight variations in HRS (see Figure. 3c and Supplementary Figure. 6a). These observations indicate the long-time stability of the device, an advantageous feature for device application, and also further confirm that the topological protection property plays an importance role in stabilizing the domain-wall conductions states, ensuring the good data retention against thermal perturbations.

Another important aspect of memory performance is endurance property, which reflects the repeatability of data restoration operation. As illustrated in Figure. 3d, the device is subjected to a set of reversible switching voltage ($V_S$) pulses (±6.0 V with pulse width of 100 μs) for up to $2×10^6$ cycles, and the conductive levels of LRS and HRS are recorded at several intervals. It is found that the device can maintain its well-established resistance switching functionality during the fatigue testing, wherein the readout current of LRS and HRS show very small variation in sequence up to $2×10^6$ cycles switching cycles (see Figure. 3d and Supplementary Figure. 8), indicating a good fatigue resistant behavior. This agrees well with the high repeatability of the *I-V* curves shown in Fig. 2, beneficial from the topological protection property.

To provide a more comprehensive understanding of the memory performance, it is essential to understand the switching dynamic behaviors of the conduction states. Figure. 3e shows the fast-switching dynamics under varying electric pulses $V_S$, wherein a constant $V_r$ ~ 2.5 V is used to read out the different resistance states (see Methods). One can see that along with the resistance state switches from HRS to LRS, an abrupt jump of readout current occurs at a characteristic time ($t_0$), which cannot not conform the typical Kolmogorov-Avrami-Ishibashi (KAI) relation [35]. As $V_S$ increases, $t_0$ shows a quick reduction. The extracted $t_0$ – $I/V_S$ curve is shown with semilogarithmic form in Figure. 3f, which can well fit to the Merz law of $t_0 = \tau_0 exp[(E_a/E)^\mu]$ [36,37], wherein $E_a$ is activation field determined to be 22.4 MV cm$^{-1}$, and $\mu$ is a corrected component of 1. It is noted that the abrupt transition from HRS to LRS is persisted as the $V_S$ decreases from 9 V to 6 V, which is unlike typical polarization switching behavior often showing a gradual switching process at low applied voltage. We therefore conjecture that the switching of the topological center state is more like a collective behavior and the whole domain state switches suddenly to another state once the driven bias is sufficient to overcome the energy barrier, leading to the abrupt readout current change with very few intermediate states.

**Individually writing and reading data bit in a device array**

Moreover, the ability of controlled writing and erasing operation in a 3×3 device array is demonstrated in Figure. 4a. It is seen that the exposed BFO regions in all the devices show uniform low conductivity at the initial state, which can also be considered as HRS. After applying a bias voltage of +6 V on eight selected devices (except one for reference), one can clearly see that the cross-shaped current patterns appear in the selected BFO regions in the C-AFM image, reflecting the formation of center-convergent domains and LRS. After that, five of the written memory units framed by the black dashed line are erased by applying a bias of -6.0 V. The cross-shaped current disappears in all the five devices, suggesting the conversion from LRS to HRS and the formation of low conductive T-T CDWs. The corresponding PFM images for the devices after being subjected to different bias voltages are shown in Supplementary Figure. 9, which verify the manipulation of center type domain states in respective devices, in accordance well with the change of conduction states. The above results well demonstrate the capability of individual data addressing, programmable writing and reading, as well as good uniformity of the devices, which make them promising for implementation in memory devices with nondestructive current readout.

In practical memory devices, these functionalities should be implemented with the use of bit lines and word lines. We therefore proposed a conceptual crossbar structure of memory devices as illustrated in Figure. 4b. The single memory unit consists of a central electrode/circular BFO area/surrounding electrode structure whereby the two electrodes are respectively attached to bit and word lines. Such device architecture with a high integration density enables the data writing and reading operations for any randomly selected individual device unit from its corresponding set of word and bit lines. To elaborate the feasibility this scheme, we also fabricated an array of ring-shaped electrodes with word lines by using electron beam lithography (EBL) technique (see Methods), which further demonstrates the ability of individual data writing and reading (see Figure. 4c & Supplementary Figure. 10). More interestingly, this structure can be expediently extended to a three-dimensional (3D) stacking structure. As shown in Figure. 4d, the central electrodes of the memory units now penetrate the two-dimensional crossbar device layers, acting as the vertical pile lines to realize vertical (Z-axis) data addressing. Such 3D stacking structure can further enhance the data storage density, which creates a new pathway for developing ultrahigh density memories based on the topological protected conductive domain wall devices.

The above results reflect promising potential for developing high density topological protected domain wall memories that exhibit fatigue resistance, and long-term retention as well as low energy consumption properties. Besides, the device array structure is also compatible with large area fabrication technique. For instance, we have successfully fabricated large-area of device array *via* a nanosphere lithography technique, and the as fabricated devices also show similar memory functionalities as mentioned above (see Supplementary Figure. 11). These beneficial features also provide a paradigm for developing new concept nano-electronic device based on the programmable functional topological domains, which may spur more efforts to explore the immense application potentials in ferroelectric topologies and eventually fulfilled the field of domain wall electronics or topological electronics.

**Conclusion**

In summary, we have successfully fabricated the well-designed and high performance memory devices based on conductive domain walls confined in center topological domains produced in high-quality epitaxial BFO film. By using a pair of delicately designed co-axial electrodes, one can deterministically create and reversibly switch the center topological domains, leading to a sharp switch between distinct domain wall conduction states. This can be implemented in nonvolatile memory devices that allow nondestructive current readout of different domain wall conduction states, with a large On/Off resistance ratio over $10^4$ and a high readout current. Moreover, owing to the topological protection nature of center domains, the devices exhibit a long retention time of $10^6$ s and a high fatigue resistance of millions of switching cycles. The devices also show the compatibility to high density integration crossbar array, which provides a new paradigm for developing topological electronic devices.

**References**

1. G. Catalan, J. Seidel, R. Ramesh, J. F. Scott, *Rev. Mod. Phys.* **2012**, 84, 119.

2. G. F. Nataf, M. Guennou, J. M. Gregg, D. Meier, J. Hlinka, E. K. H. Salje, J. Kreisel, *Nat. Rev. Phys.* **2020**, 2, 634.

3. J. Seidel, R. K. Vasuevan, N. Valanoor, *Adv. Electron. Mater.* **2016**, 2, 1500292.

4. L. Li, L. Xie, X. -Q. Pan, *Rep. Prog. Phys.* **2019**, 82, 126502.

5. P. S. Bednyakov, B. I. Sturman, T. Sluka, A. K. Tagantsev, P. V. Yudin, *Npj Comput. Mater.* **2018**, 4, 65.

6. J. Seidel, L. W. Martin, Q. He, Q. Zhan, Y.-H. Chu, A. Rother, M. E. Hawkridge, P. Maksymovych, P. Yu, M. Gajek, N. Balke, S. V. Kalinin, S. Gemming, F. Wang, G. Catalan, J. F. Scott, N. A. Spaldin, J. Orenstein, R. Ramesh, *Nat. Mater.* **2009**, 8, 229.

7. J. Seidel, D. Fu, S.-Y. Yang, E. Alarco´n-Llado´, J. Q. Wu, R. Ramesh, J. W. Ager III, *Phys. Rev. Lett.* **2011**, 107, 126805.

8. Q. He, C.-H. Yeh, J.-C. Yang, G. Singh-Bhalla, C.-W. Liang, P.-W. Chiu, G. Catalan, L. W. Martin, Y.-H. Chu, J. F. Scott, R. Ramesh, *Phys. Rev. Lett.* **2012**, 108, 067203.

9. A. Crassous, T. Sluka, A. K. Tagantsev, N. Setter. *Nat. Nanotech.* 2015, 10, 614.

10. J. Jiang, L. B. Zi, H. C. Zhi, H. Long, D. W. Zhang, Q. H. Zhang, A. S. Jin, H. P. Min, J. F. Scott, C. S. Hwang, A.-Q. Jiang, *Nat. Mater.* **2018**, 17, 49.

11. T. Sluka, A. K. Tagantsev, P. Bednyakov, N Setter, *Nat. Commun.* 2013, 4, 1808.

12. J. Guyonnet, I. Gaponenko, S. Gariglio, P. Paruch, *Adv. Mater.* **2011**, 23, 5377.

13. M. Schroder, A. Haußmann, A. Thiessen, E. Soergel, T. Woike, L. M. Eng, *Adv. Funct. Mater.* **2012**, 22, 3936.

14. W. D. Wu, Y. Horibe, N. Lee, S.-W. Cheong, J. R. Guest, *Phys. Rev. Lett.* **2012**, 108, 077203.

15. P. Sharma, Q. Zhang, D. Sando, C. H. Lei, Y. Y Liu, J. Y. Li, V. Nagarajan, J. Seidel, *Sci. Adv.* **2017**, 3, e1700512.

16. G. Tian, W. D. Yang, X. Song, D. F. Zheng, L. Y. Zhang, C. Chen, P. L. Li, H. Fan, J. X. Yao, D. Y. Chen, Z. Fan, Z. P. Hou, Z. Zhang, S. J. Wu, M, Zeng, X. S. Gao, J. -M. Liu, *Adv. Funct. Mater.* **2019**, 29, 1807276.

17. A. Q. Jiang, W. P. Geng, P. Lv, J.-W. Hong, J. Jiang, C. Wang, X. J. Cai, J. W. Lian, Y. Zhang, R. Huang, W. Z. David, J. F. Scott, C. S. Hwang, *Nat. Mater.* **2020**, 19, 1188.

18. J. P. V. McConville, H. D. Lu, B. Wang, Y. Tan, C. Cochard, M. Conroy, K. Moore, A. Harvey, U. Bangert. L.-Q. Chen, A. Gruverman, J. M. Gregg, *Adv. Funct. Mater.* **2020**. 30, 2000109.

19. D. Y. Chen, Z. H. Chen, Q. He, J. D. Clarkson C. R. Serrao, A. K. Yadav, M. E. Nowakowski, Z. Fan, L. You, X. S. Gao, D. C. Zeng L. Chen, A. Y. Borisevich, S. Salahuddin, J. -M. Liu, J. Bokor, *Nano Lett.* **2017**, 17, 486.

20. R. K. Vasudevan, A. N. Morozovska，E. A. Eliseev，J. Britson，J. C. Yang，Y. H. Chu，P. Maksymovych，L. Q. Chen，V. Nagarajan，S. V. Kalinin, *Nano Lett.* **2012**, 12, 5524.

21. G. Tian, W. D. Yang, D. Y. Chen, Z. Fan, Z. P. Hou, M. Alexe, X. S. Gao, *Nat. Sci. Rev.* **2019**, 6, 684.

22. J. Seidel, *Nat. Mater.* **2019**, 18, 188.

23. S. Q. Chen, S. Yuan, Z. P. Hou, Y. L. Tang. J. P. Zhang, T. Wang, K. Li, W. W. Zhao, X. J. Liu, L. Chen, *Adv. Mater.* **2021**, 33, 2000857.

24. Y. L. Tang, Y. L. Zhu, X. L. Ma, A. Y. Borisevich, A. N. Morozovska, E. A. Eliseev, W. Y. Wang, Y. J. Wang, Y. B. Xu, Z. D. Zhang, S. J. Pennycook, *Science.* **2015**, 348, 547.

25. A. K. Yadav, C. T. Nelson, S. L. Hsu, Z. Hong, J. D. Clarkson, C. M. Schleputz, A. R. Damodaran, P. Shafer, E. Arenholz, L. R. Dedon, D. Chen, A, Vishwanath, A. M. Minor, L. Q. Chen, J. F. Scott, L. W. Martin, R. Ramesh, *Nature.* **2016**, 530, 198.

26. S. Das, Y. L. Tang, Z. Hong, M. A. P. Goncalves, M. R. McCarter, C. Klewe, K. X. Nguyen, F. Gomez-Ortiz, P. Shafer, E. Arenholz, V. A. Stoica, S.-L. Hsu, B. Wang, C. Ophus, J. F. Liu, C. T. Nelson, S. Saremi, B. Prasad, A. B. Mei, D. G. Schlom, J. Íñiguez, P. García-Fernández, D. A. Muller, L. Q. Chen, J. Junquera, L. W. Martin, R. Ramesh, *Nature.* **2019**, 568, 368.

27. Z. W. Li, Y. J. Wang, G. Tian, P. L. Li, L. N. Zhao, F. Y. Zhang, J. X. Yao, H. Fan, X.

Song, D. Y. Chen, Z. Fan, M. H. Qin, M. Zeng, Z. Zhang, X. B. Lu, S. J. Hu, C. H. Lei, Q. F. Zhu, J. Y. Li, X. S. Gao, J. -M. Liu, *Sci. Adv.* **2017**, 3, e1700919.

28. J. Ma, J. Ma, Q. H. Zhang, R. C. Peng, J. Wang, C. Liu, M. Wang, N. Li, M. F. Ghen, X. X. Cheng, P. Gao, L. Gu, L. -Q. Chen, P. Yu, J. X. Zhang, C. -W. Nan, *Nat. Nanotech.* **2018**, 13, 947.

29. W. D. Yang, G. Tian, Y. Zhang, F. Xue, D. F. Zheng, L. Y. Zhang, Y. D. Wang, C. Chen, Z. Fan, Z. P. Hou, D. Y. Chen, J. W. Gao, M. Zeng, M. H. Qin, L. -Q. Chen, X. S. Gao, J. -M. Liu, *Nat. Common.* **2021**, 12, 1306.

30. W. Peng, J. Mun, Q. Xie, J. S. Chen, L. f. Wang, M. Kim, T. W. Noh, *npj Quantum Mater.* **2021**, 6, 48.

31. J. Kim, M. You, K. E. Kim, K. Chu, C. -H. Yang, *npj Quantum Mater.* **2019**, 4, 29.

32. J. B. Johnson, *Phys. Rev.* **1928**, 32, 97.

33. H. Nyquist, *Phys. Rev.* **1928**, 32, 110.

34. E. Linn, R. Rosezin, Carsten. Kuegeler, R. Waser, *Nat. Mater.* **2010**, 9, 403.

35. Y. W. So, D. J. Kim, T. W. Noh, J.-G. Yoon, T. K. Song, *Appl. Phys. Lett.* **2005**, 86, 092905.

36. W. J. Merz, *Phys. Rev.* **1954**, 95, 690.

37. T. Tybell, P. Paruch, T. Giamarchi J.-M. Triscone, *Phys. Rev. Lett.* **2002**, 89, 097601.

**Acknowledgments**

The authors would like to acknowledge the financial support from the National Key Research and Development Programs of China (Grant Nos. 2016YFA0201002, 2016YFA0300101), the National Natural Science Foundation of China (Grant Nos. 11674108, 51272078, 52002134), the Science and Technology Program of Guangzhou (No. 2019050001), the project for Basic and Applied Basic research Foundation of Guangdong Province (No.2019A1515110707), the Natural Science Foundation of Guangdong Province (No. 2016A030308019), the Science and Technology Planning Project of Guangdong Province (No. 2019KQNCX028), and the Natural Science Foundation of South China Normal University (No. 19KJ01).

**Author contributions**

X. S. Gao conceived and designed the experiments. W. D. Yang conducted the main experiments. G. Tian, H. Fan, Y. Zhao and H. Y. Chen contributed to the sample fabrication and XRD measurements. W. D. Yang carried out the AFM, PFM, C-AFM measurements. G. Tian, D. Y. Chen, Z. Fan, Z. P. Hou, J. W. Gao contributed to the data interpretation. X. S. Gao, and J.-M. Liu conducted the data interpretation and co-wrote the article. All authors discussed the results and commented on the manuscript.

**Competing interests**

The authors declare no competing financial interest.

**Data availability**

The data sets that support the findings of this study are available from the corresponding authors on reasonable request.

**Methods**

**Fabrication of the devices.** The BiFeO$_3$ thin film was deposited on the (001)-oriented SrTiO$_3$ substrate by pulsed laser deposition (PLD) using a KrF excimer laser (wavelength $\lambda$ = 248 nm) at a substrate temperature of 660 ℃ in an oxygen ambient of 6 Pa. The laser pulse energy was 300 mJ with a repetition rate of 8 Hz. Then an Au layer with thickness of 10 nm was deposited on the BFO film surface via thermal evaporation. To fabricate the device, the AFM tip was used to scratch away a circular shaped Au region with a constant tip pressure of 40 μN and velocity of 1 μm s$^{-1}$, as schematic illustrated in Supplementary Figure. S2. Consequently, a circular exposed region of BFO film surrounded by the Au electrode is formed. Moreover, such device can also be fabricated via an electron beam lithography (EBL) on a layer of polymethyl methacrylate resist, which is followed by the deposition of Au layer and a final lift-off process.

**Microstructural characterizations.** The epitaxial structures of BFO film were characterized by X-ray diffraction (PANalytical X′Pert PRO), including $\theta$-$2\theta$ scanning and reciprocal space mapping (RSM) along the (203) diffraction spot. The topography images were taken by atomic force microscopy (AFM) (Cypher, Asylum Research).

**PFM and C-AFM characterizations.** The ferroelectric domain structures of the devices were characterized by piezoresponse force microscopy (PFM) (Cypher, Asylum Research) using conductive PFM probes (Arrow EFM, Nanoworld). The local piezoresponse loop measurements were carried out by fixing the PFM probe on a randomly selected BFO region and then applying a triangle square waveform accompanied with a small *ac* driving voltage from the probe. Using vector PFM mode, one can simultaneously map the vertical and lateral piezoresponse signals of the exposed BFO area. To determine the domain structures, both the vertical and lateral PFM images were recorded at 0° and 90° sample rotation angles. The current distribution maps were characterized by conductive atomic force microscopy (C-AFM) using conductive probes (HA_HR_DCP, TipNano).

**Electrical characterization.** The *I-V* curves were measured using a precise current meter (Keithley 6430) in voltage-sweep mode, which was connected to the conductive AFM tip that was fixed on the center of the exposed BFO region serving as a stationary electrode. The memory performance tests were conducted via C-AFM mode, wherein a sequence of electric pulses with triangularly variational amplitudes were applied on the Au layer to trigger the domain switching with a pulse width of 5 ms. Immediately after each pulse, the conduction state was read by using a fixed read voltage ($V_r$) of 2.5 V. For the endurance testing, we applied reversible electric pulses (with square pulse width of 100 μs, and maximum voltage 6.0 V) for up to $2\times10^6$ switching cycles. To determine the resistive switching time, we applied square electric pulses ($V_s$) for varying durations by using a pulse generator (Agilent 33250A), and then record the conduction state by a precision current meter (Keithley 6430) at a read bias of 2.5V after each pulse, and the switching time was determined as the pulse duration during which the resistive change from HRS to LRS is completed. It is also noted that, to access the small device, both the pulse generator and precision current meter were connected with the conductive AFM tip.

**Figures**

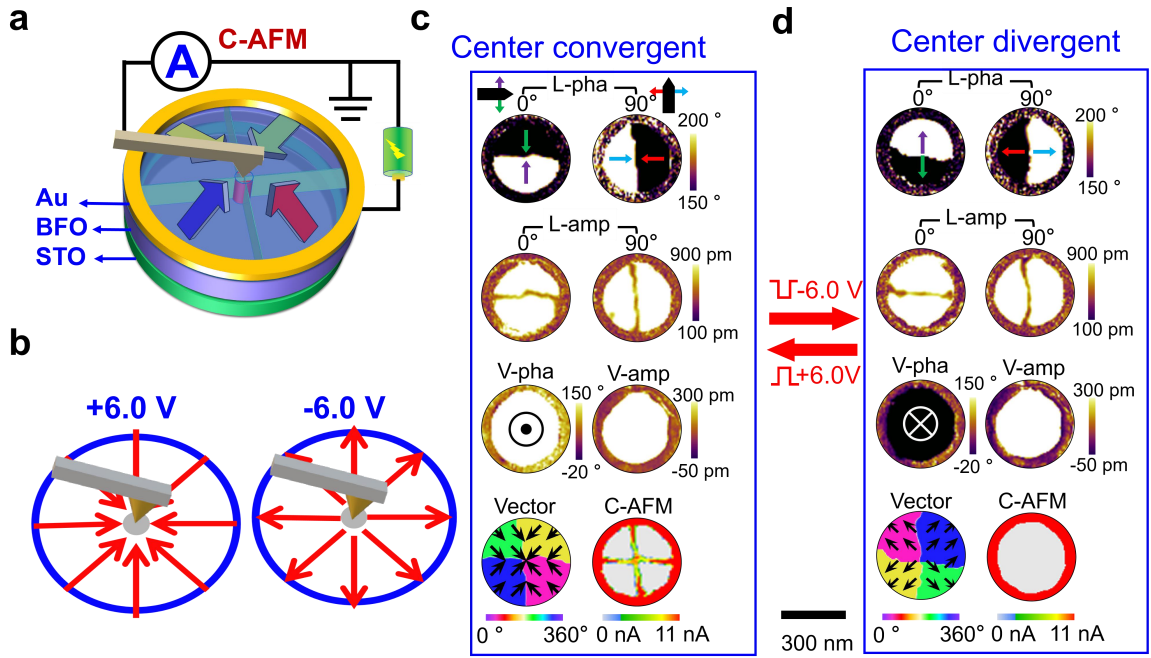

**Figure. 1 Creation of quadrant center topological domains. a** The schematic diagram of the device structure. It is noted the bias voltage were applied on the Au electrode and the tip was grounded in all the figures. **b** The schematic illustrations of radial electric fields on circular exposed BFO region generated by applying electric bias between the tip and Au electrodes. **c,d** The created quadrant center convergent and divergent domain states along with their corresponding domain wall conduction states: vector PFM images and corresponding in-plane polarization vector maps, as well as CAFM maps for both center convergent (c) and divergent states (d), which are obtained after applying tip biases of +6 V and -6 V, respectively. To construct the in-plane vector maps, lateral PFM amplitude (L-Amp) and phase (L-Pha) were recorded for sample rotation at 0° and 90°.

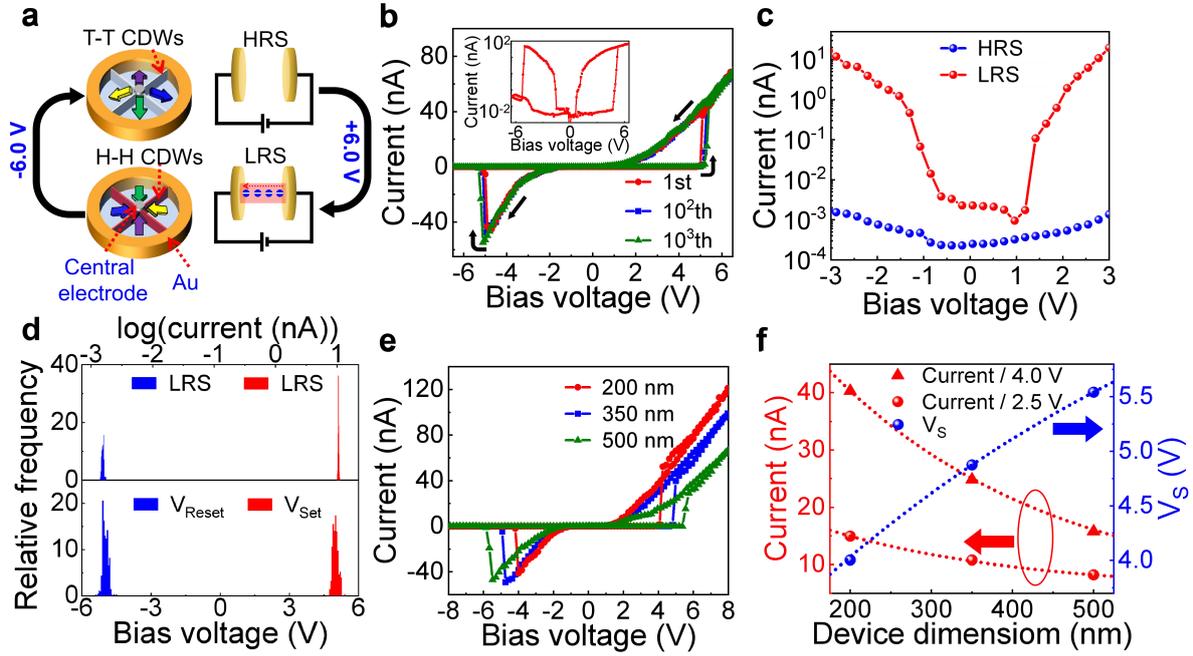

**Figure. 2 Resistance switching behaviors of the devices. a** Schematic working principle based on creation and erasure of conductive domain wall states. **b** Quasi-static *I-V* curves for 1,000 consequent switching cycles at a smaller bias range of ±7 V. Inset shows a semi-logarithmic *I-V* curve. **c** *I-V* curves at a smaller bias range of ±3 V for both *HRS* and *LRS*. **d** Distributions of *log(current)* for both *LRS/HRS* and switching voltages (set and reset voltages: $V_{set}$ and $V_{reset}$) derived from the statistics on $10^3$ consecutive *I-V* cycles (shown in **a**). **e, f** Resistive switching characteristics for different device dimension (i.e. diameters of the circular exposed BFO regions): *I-V* hysteresis curves for devices with different characteristic device dimension from 200 nm to 500 nm **(e)**, and length dependent readout currents (for two different Vr) and switching voltages (set voltage) extracted from the *I-V* curves **(f)**.

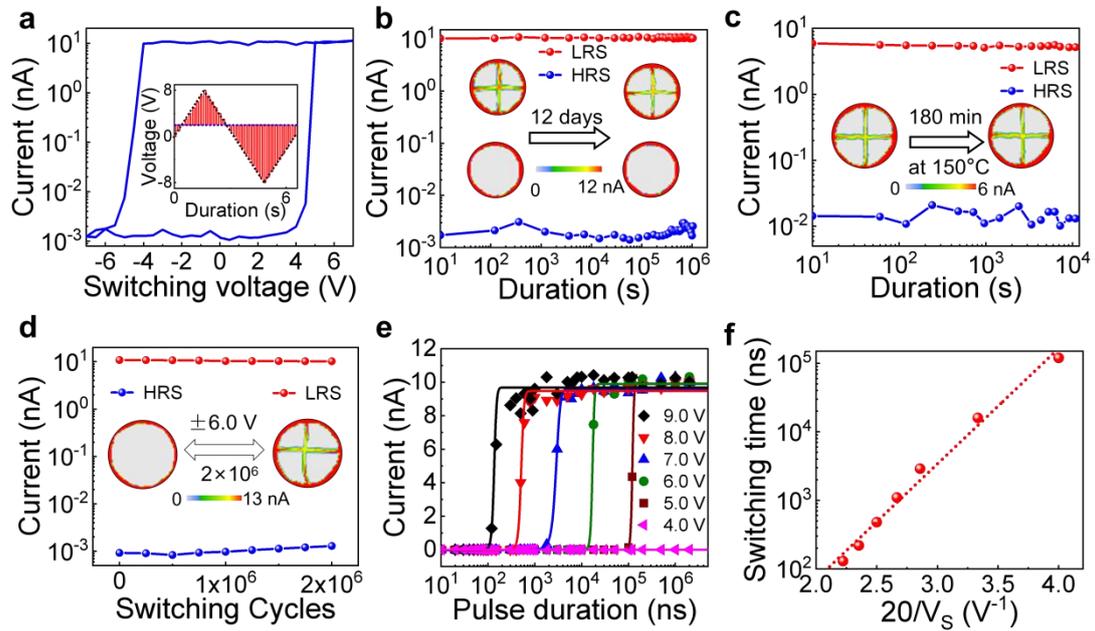

**Figure. 3 Performance of the devices. a** The readout current *versus* pulse voltage obtained after applying different write pulses (at a fix reading bias of 2.5 V), indicating the memory switching window. **b,c** The retention properties for *LRS* and *HRS* at room temperature of 25 °C (b) and at an elevated temperature of 150 °C (c). **d** Endurance properties for both *LRS* and *HRS* over 2×10⁶ switching cycles. **e** Evolutions of readout current with respect to pulse duration under different pulse voltages. **f** Relationship between switching time and applied pulse voltage extracted from (e), indicating a minimum switching time of 120 ns at a pulse voltage of 9.0 V.

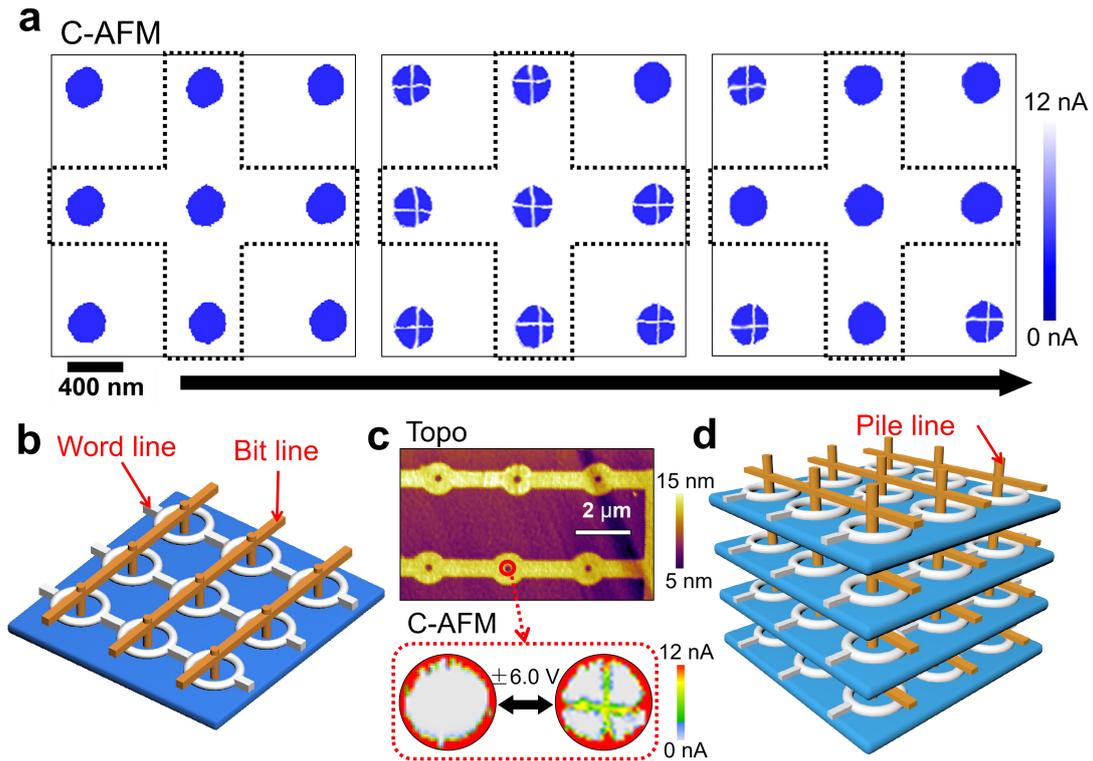

**Figure. 4 Writing and erasing of individual data bit in a device array and schematic conceptual crossbar integrated devices. a** Individually writing and erasing of data bit in a 3×3 device array. Here the Au layer is grounded, and the AFM tip acts as a movable electrode to address the individual device with ±6.0 V bias voltage. **b** A conceptual crossbar-integrated domain wall memory device. **c** Example of a device array connected by word lines fabricated by EBL technique, which shows the capability of individually data writing and erasing process. **d** Conceptual three-dimensional integrated memory device with stacking structure, wherein the central electrodes are connected by the pile lines.

**Supplementary Information**

**Topologically protected ferroelectric domain wall memory with large readout current**

Wenda Yang,[1] Guo Tian,[1,*] Hua Fan,[2] Yue Zhao[2], Hongying Chen,[1] Luyong Zhang,[1] Yadong Wang,[1] Zhen Fan,[1,*] Zhipeng Hou,[1] Deyang Chen,[1] Jinwei Gao,[1] Min Zeng,[1] Xubing Lu,[1] Minghui Qin,[1] Xingsen Gao,[1*] and Jun-Ming Liu[1,3]

*Corresponding authors:
X. S. Gao, xingsengao@scnu.edu.cn,
Z. Fan, fanzhen@m.scnu.edu.cn,
G. Tian, guotian@m.scnu.edu.cn

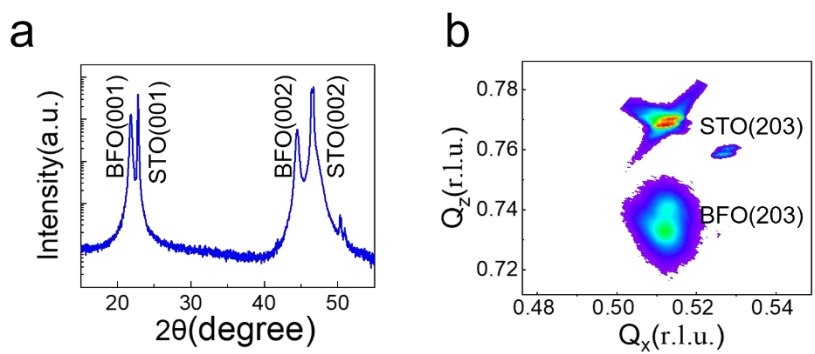

**Fig. S1 Structure characteristics of BFO / STO thin film.** (a) X-ray diffraction (XRD) *θ - 2θ* scan and (b) (203) Reciprocal space mapping (RSM) of the BFO thin film grown on the (001) STO substrate. The XRD *θ - 2θ* diffraction pattern exhibits only (00*l*) diffraction peaks of BFO and STO, excluding the existence of impurity phases. Besides, the diffraction pattern of BFO in the (203) RSM map show two spots, indicating that the BFO unit cell shows slightly monoclinic tilt. Based on XRD and RSM results, the lattice constants of BFO are determined as a = b ~ 3.905 nm and c ~ 4.07 nm.

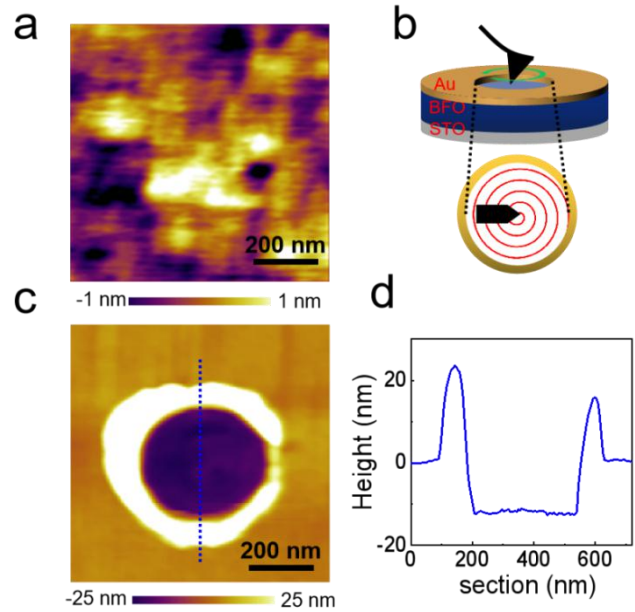

**Fig. S2 Fabrication of circular exposed BFO region via AFM lithography.** (a) The AFM topological image of as-grown Au/BFO/STO film. (b) Schematic diagram shows the lithography process for fabricating circular exposed BFO area, by scratching away the Au region with a moving AFM tip that moved in a spiral trajectory with a tip pressure of 40 μN. The resolution of the created exposed BFO area was limited by both the size of the AFM tip and the precision of the tip motion, which show a rather high resolution within 20 nm. (c) AFM morphology of the circular exposed BFO region. (d) Surface height profile along the dashed line extracted from the AFM image in Fig. S2c, showing that the exposed BFO area had a diameter of about 350 nm. The height difference between the exposed BFO area and the surrounding Au covered area (except edge region) was 12 nm, which is exactly the thickness of Au layer.

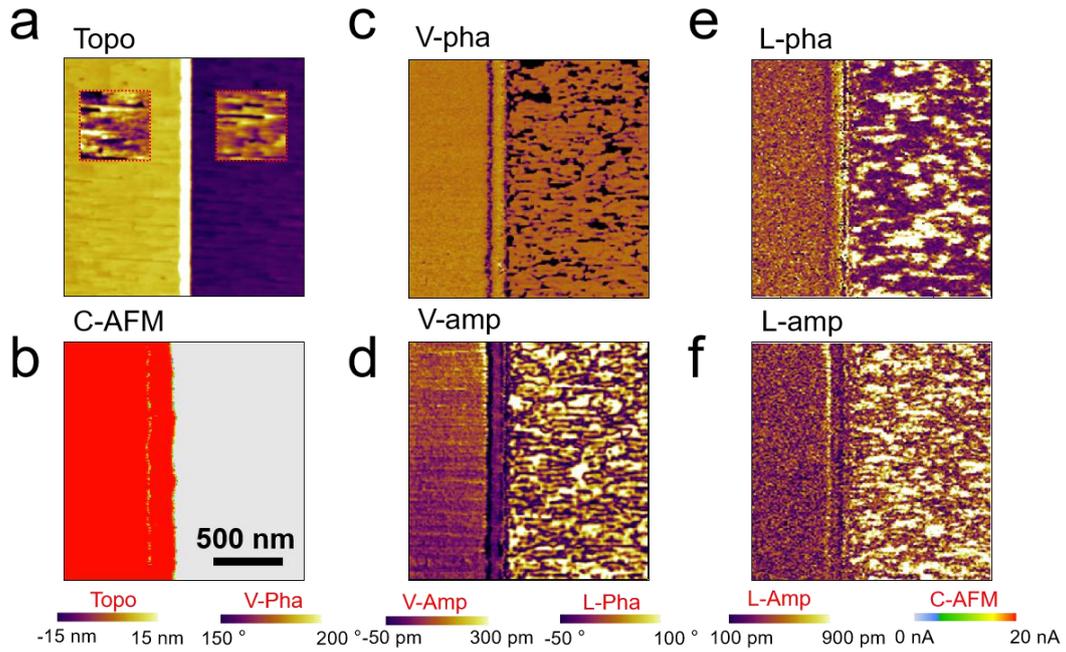

**Fig. S3 Domain structure and conductivity of exposed BFO area.** (a) Topography of the Au/BFO/STO heterostructure where the right half of the Au layer was scraped off by the AFM tip, leaving an exposed BFO area. The smooth surface morphology reflects that the scraping process caused neglectable damage to the BFO layer. (b) Corresponding C-AFM images under a sample bias of 2.5 V. The Au covered region exhibits high conductive current compared with that of the exposed BFO area. (c-f) PFM images of the sample, including vertical PFM phase (V-Pha) and amplitude (V-Amp) images (c, d), lateral PFM phase (L-Pha) and amplitude (L-Amp) images (e, f). The lateral PFM images of the exposed region indicate the mosaic domain pattern of as-grown BFO film underneath the Au layer.

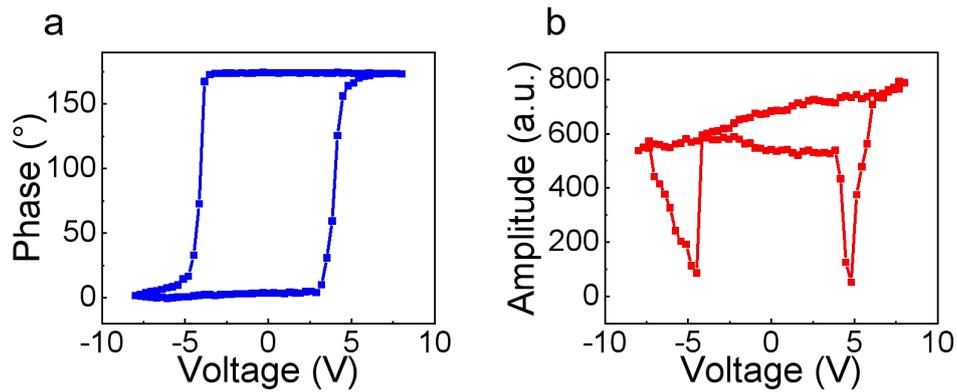

**Fig. S4 Piezoresponse hysteresis loops of the exposed BFO region.** (a) The square phase-voltage hysteresis loop, and (b) The butterfly-like amplitude-voltage loop. The measured hysteresis loops of the exposed BFO area (including both phase-voltage hysteresis and butterfly-like amplitude- oltage loop) are similar to those of BFO film with bottom electrodes. One can clearly see that the coercive voltage is about ±4.8 V.

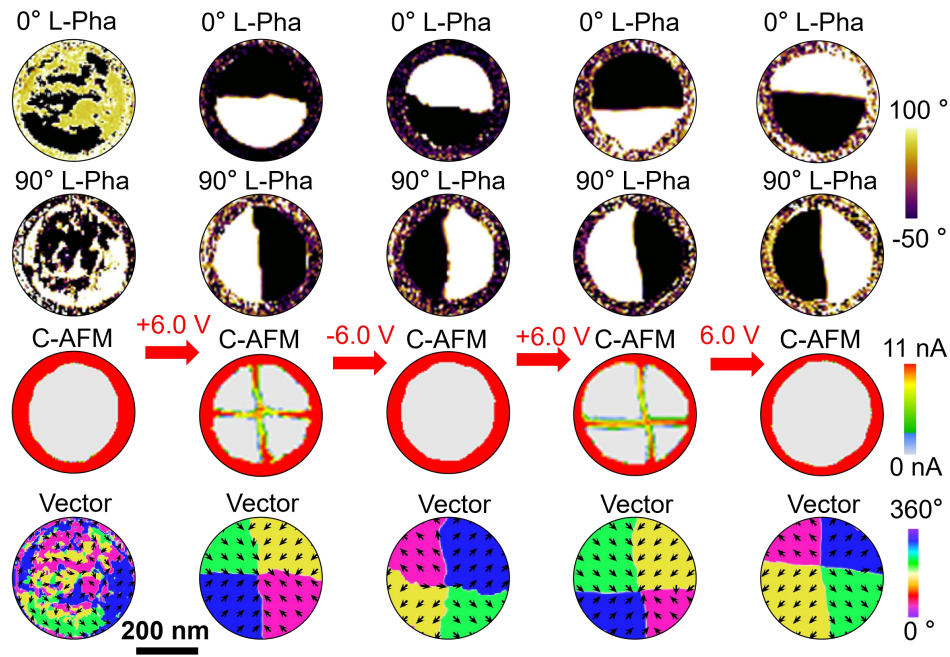

**Fig. S5 Center domain structure creation and reversible switching triggered by pulse voltages of ±6.0 V.** From left column to right column shows the domain structure for initial

state, and those obtained after applying pulse voltages of +6 V, -6 V, +6 V and -6 V sequentially. From top line to bottom lines presents lateral PFM phase images recorded at sample rotations of 0° and 90° and corresponding C-AFM images, as well as reconstructed piezoresponse vector map. One can clearly see that after being subjected to bias voltage of +6.0 V, the initially non-conductive mosaic domain structure converts to a center-convergent state with conductive H-H CDWs as reflected by the cross-shaped current pattern in C-AFM images. Applying -6 V and +6 V can reversibly switch the center-type domains between a convergent state (with conductive H-H CDWs) and a divergent state (with much less conductive T-T CDWs), leading to a resistive switching behavior.

A. **Conduction mechanism of the H-H CDWs**

To gain further insight into the conduction mechanism of H-H CDWs, we conducted a temperature-dependent C-AFM measurement. As shown in Fig. S6a, the currents in H-H CDWs exhibit gradual decrease with temperature increasing from 25 °C to 150 °C. Besides, despite the current decay, the cross-shaped current pattern can still be observed up to 150 °C. The currents can be fully restored to the initial values when the sample is cooled back to 25 °C, indicating the good stability of the topological domain structure against high temperature. From the temperature-dependent C-AFM images, we extracted the current profiles shown in Fig. S6b, and accordingly plotted the current-temperature *(I-T)* curve (see Fig. S6c). The H-H CDWs exhibit a negative temperature coefficient of current, which fits to the model of metallic conduction relation: $I \sim I_0(1 + a(T - T_0))^{-1}$, demonstrating the metallic nature of H-H CDWs [1,2]. The metallic conducting behavior can be understood by the charge accumulation. Inside the H-H CDWs, the high-density positive bound charges need to be screened by negative charges. When applying the tip bias, the injected electrons can be attracted by H-H CDWs to compensate for the positive bound charges, thus generating a local conductive path.

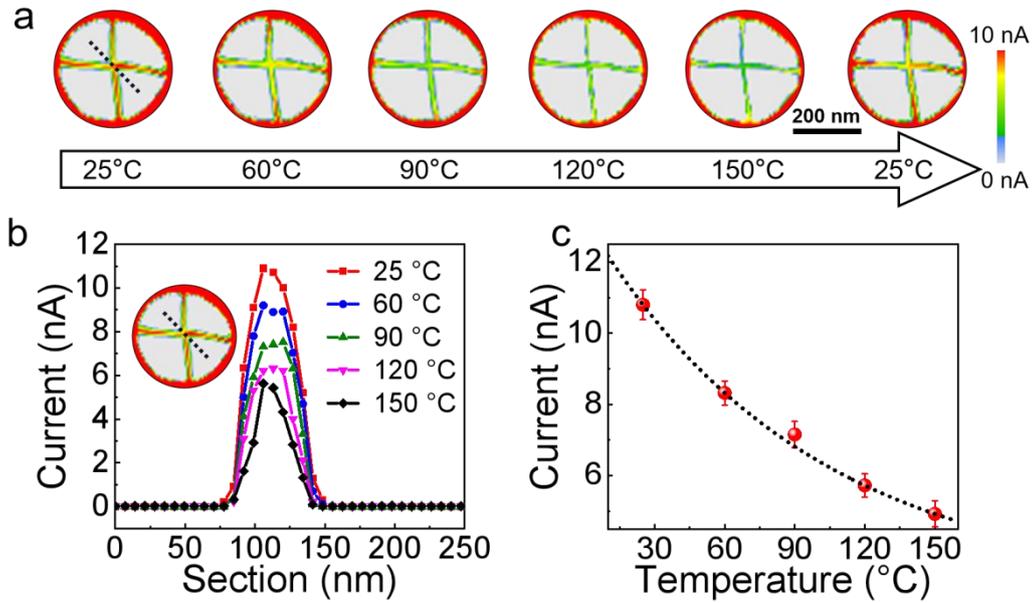

**Fig. S6 Temperature dependent conduction behaviors for the H-H CDWs.** (a) The C-AFM images of the center-convergent domain state recorded at different temperatures. (b) Extracted current profiles from the C-AFM images along the dashed line recorded at different temperatures. (c) Temperature dependent conductive current (*I-T*) curve for the center-convergent domain state, which can fits well to the well-known metallic conductance relation $I \sim I_0(1 + a(T - T_0)^{-1})$.

## B. The effect of device dimension on read-out current

We examined the effect of characteristic dimension of device (i.e. diameter of the exposed BFO region) on the read-out current. As shown in Fig. S7, the devices with length of 200 nm, 350 nm and 500 nm respectively show similar resistive switching behavior, as reflected by appear and disappear of the hallmark of the cross-shaped conducting paths in C-AFM images before and after application of +6.0 V tip bias. From the current profiles extracted form corresponding C-AFM images, the readout current at a fixed reading bias of 2.5 V shows an appearance increase with length shrinkage. Predictably, benefiting from the advances of fabrication techniques, further size shrinkage is expected to improve the performance in both storage density and On/Off ratio.

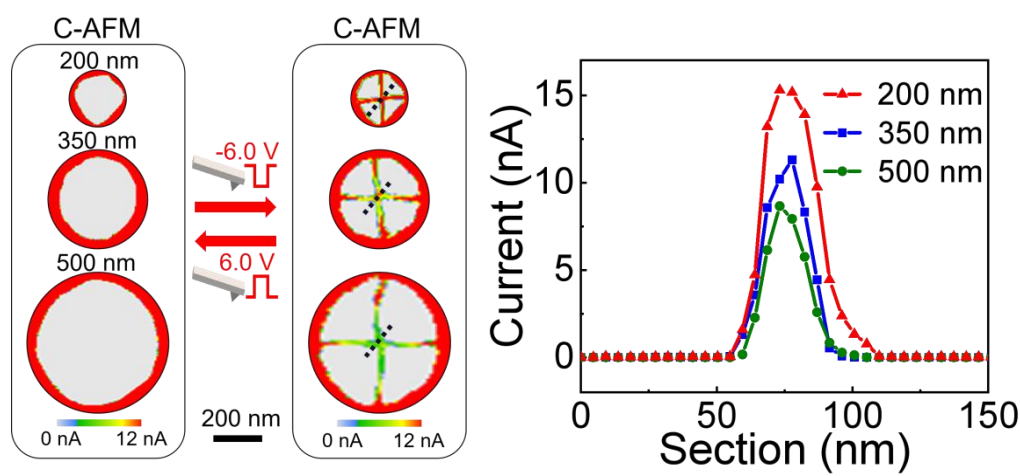

**Fig. S7 Read-out currents in LRS for the devices with different characteristic dimension.** (a,b) The CAFM maps indicating reversely switching between center convergent (a) and divergent (b) states for different characteristic dimension of 200 nm, 350nm and 500nm respectively. All of them exhibit switchable conductive levels between HRS and LRS switched by tip bias of ±6.0 V. (c) Extracted current profiles from the C-AFM images in (a and b) as a function of characteristic dimension for the LRS read-out current.

## C. The fatigue properties with elongated switching cycles

We examined the fatigue properties of the device via resistance switching testing with switching cycles up to $10^6$ (Fig. S8). For this, a conductive tip was fixed on the device center and applied periodic electric pulses (with square pulse width of 100 μs, and maximum voltage ±6.0 V) up to $2\times10^6$ cycles and then collected the piezoresponse loops and current maps at different intervals. As shown in Fig. S8, after suffering $2\times10^6$ switching cycles, the device retains the ability of reversible switching between LRS and HRS, despite a small loss of read-out current for LRS.

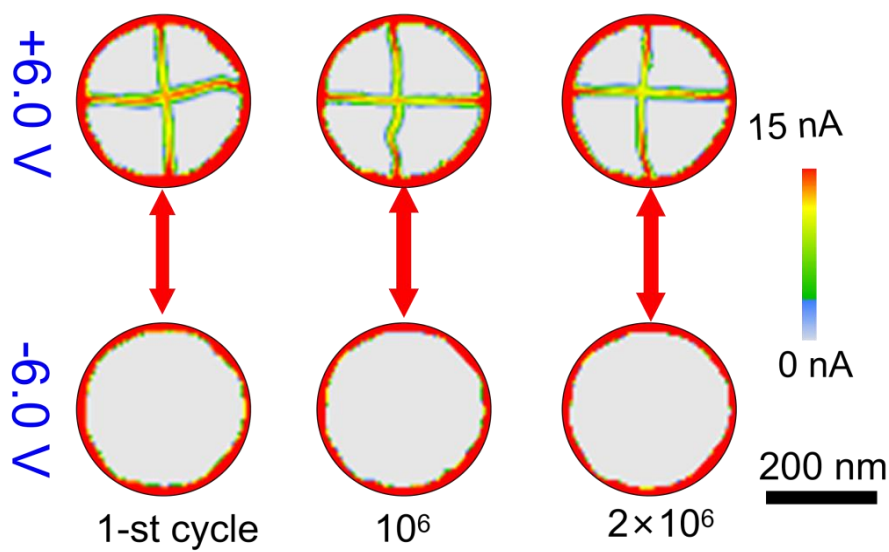

**Fig. S8 Fatigue behaviors of the device:** C-AFM maps illustrating the evolution of domain wall conduction states for both LRS and HRS after being subjected to different switching cycles.

D. **Individual writing and erasing of data bit in a device array**

The ability of independent addressing is demonstrated by individually achieving controllable conductive manipulation of single device in a 3×3 array. Nine BFO disks with uniform diameter of 350 nm were fabricated by the AFM lithography and each of them can be regarded as an independent memory unit when the movable AFM tip situates at its center. As shown in Fig. S9, the pristine BFO disks exhibit initially mosaic domain structure. Subsequently, eight units were successively excited by -6.0 V tip bias to enter center-convergent state (the last one maintains its initial state to act as a reference). One can clearly see the half-bright and half-dark contrast in both 0° and 90° PFM phase images of the modulated units, indicating the formation of center-convergent domain structure. Then five of the written units enclosed by the red line were subjected to +6.0 V tip bias, resulting in the conversion of domain structure to the center-divergent type, as demonstrated by the totally reversed contrast in PFM images. Note that the independent unit manipulation can be further realized by word lines and bit lines, satisfying the basic requirement of non-destructive readout memory devices.

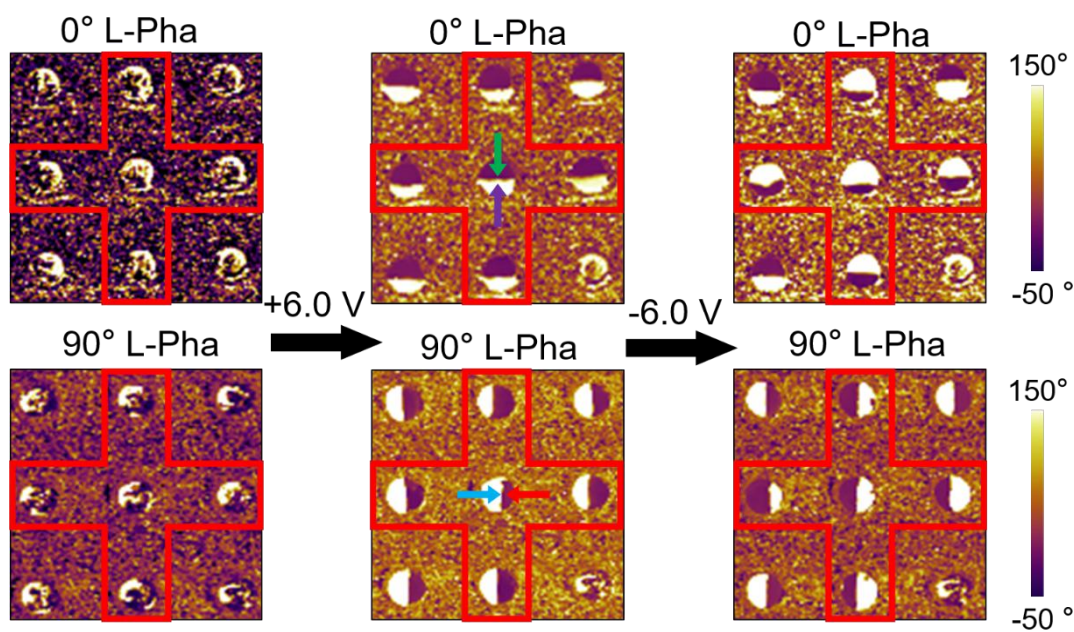

**Fig. S9 Individual creation and switching of center-type topological domain in a device**

**array.** Both 0° and 90° PFM phase images of written data bits in a 3×3 units array, after being subjected to -6.0 V and +6.0 V tip voltages.

### E. Device array fabricated by electron beam lithography (EBL)

To prove the conceptual device structure toward practical application, we fabricated an array of devices connected by word lines by using electron beam lithography (EBL) technique (see Fig. S10a). The diameter of the exposed BFO region inside each unit is about 300 nm. The C-AFM image shown in Fig. S10b indicates good conductivity of the Au layer (word lines). When the word lines are grounded, the AFM tip can individually address the different devices. The domain structures and corresponding conductive states switched by tip bias of ±6.0 V are shown in Fig. S10c, demonstrating good functionalities of devices similar to those fabricated by AFM lithography.

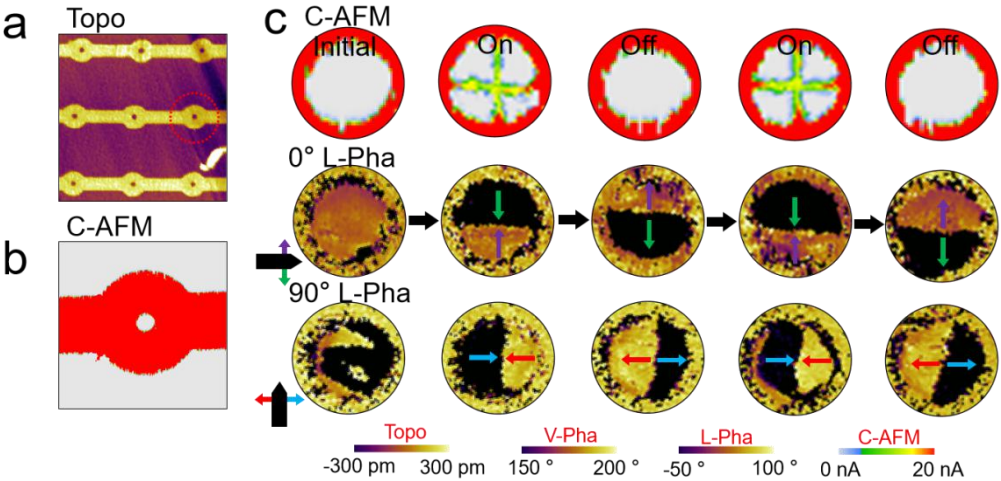

**Fig. S10 An array of device units with word lines fabricated by using electron beam lithography technique.** (a) Topography of the devices arrays. The circular electrodes as well

as the word lines are fabricated via electrode beam lithography (EBL) technique. (b) C-AFM image for a selected device which shows high conductive current of the electrode. (c) The writing and erasing of an individual device illustrated by CAFM and PFM images.

**F. Fabrication of device array via a nanosphere assisted lithography**

In addition, the device structure can be well compatible with different fabrication techniques. Shown as an example in Fig. S11, we prepared a large area device array via nanosphere lithography with polystyrene sphere (PS) arrays as templates. The detailed fabrication process is shown by the schematic in Fig. S11a. Firstly, the well-packed monolayer PS spheres were transferred onto the BFO film surface. which were subsequently subjected to size shrinkages by oxygen plasma to develop discrete ordered array. Then the Au layer with thickness of 10 nm was covered to the film surface through thermal evaporation and finally the PS mask was wiped off by chloroformic solution, leaving the exposed BFO disks surrounded by Au layer. The morphology of fabricated BFO disks is shown in Fig. S11b. The diameter of such BFO disks is about 300 nm, while can be further scaled-down to smaller dimension. The device functionality is demonstrated in Fig. S11c, which is analogue to that of the device shown in the manuscript.

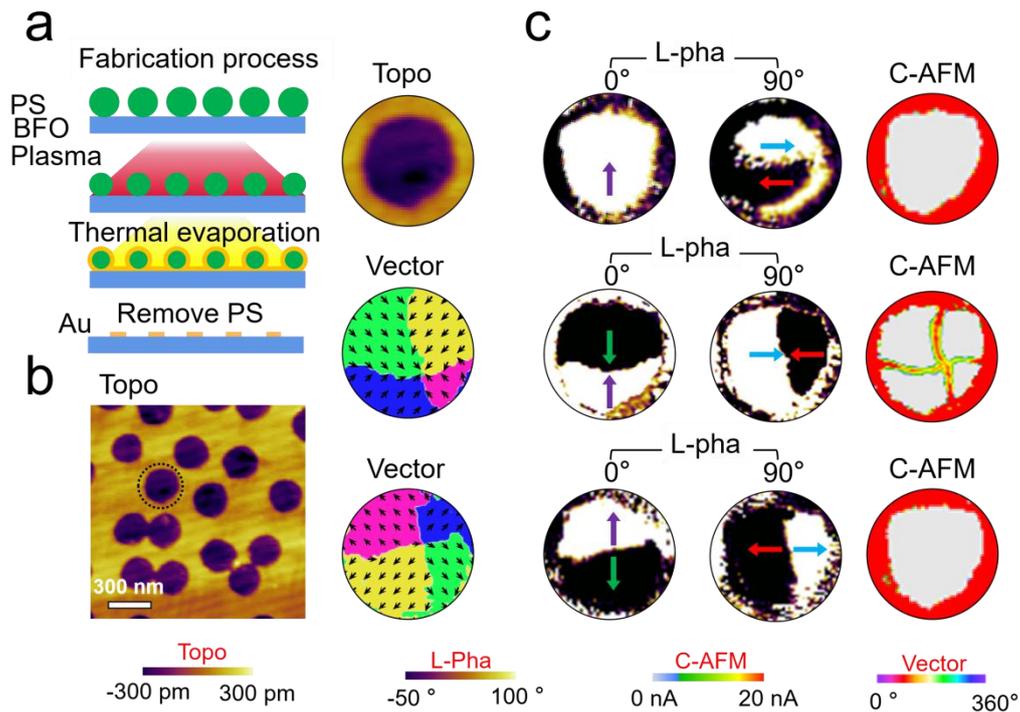

**Fig. S11 Fabrication of large-area BFO device array.** (a) Schematic fabrication process for the nanosphere lithography method. (b) The topography of a fabricated device. (c) The domain structure and conduction patterns indicating the reversible switching of center topological states and their corresponding conduction patterns.

**References:**

1. Yang, W. D. et al. Quasi-one-dimensional metallic conduction channels in exotic ferroelectric topological defects. *Nat. Common.* **12,** 1306 (2021).

2. Crassous, A., Sluka, T., Tagantsev, A. K. & Setter, N. Polarization charge as a reconfigurable quasi-dopant in ferroelectric thin films. *Nat. Nanotech.* **10,** 614-618 (2015).